%% file: 00_main.tex
\documentclass{ifacconf}

\usepackage{graphicx}      
\usepackage{natbib}        

\usepackage{listings,multicol} 
\usepackage[hyphens]{url} 
\newcommand{\myurl}[1]{\url{#1}}

\usepackage{flushend} 

\usepackage{amssymb} 
\usepackage[dvipsnames]{color}
\definecolor{mygreen}{rgb}{0, 0.7, 0}
\definecolor{myorange}{rgb}{0.9, 0.6, 0}
\definecolor{myred}{rgb}{0.8, 0, 0.2}
\newcommand{\myyes}[0]{{\color{mygreen}\checkmark}}
\newcommand{\mypartial}[0]{{\color{myorange}(\checkmark)}}
\newcommand{\myno}[0]{{\color{myred}x}}

\usepackage{rotating}

\newcommand{\mycite}[2]{\cite{#1} (#2)} 
\newcommand{\mycitep}[2]{(\cite{#1}, #2)} 

\begin{document}
\begin{frontmatter}

\title{Timeseries on IIoT Platforms: Requirements and Survey for Digital Twins in Process Industry
\thanksref{footnoteinfo}}

\thanks[footnoteinfo]{
This work has been supported by CAPRI project, which has received funding from the European Union's Horizon 2020 research and innovation programme under grant agreement No. 870062, and
s-X-AIPI project, which has received funding from the European Union's Horizon Europe research and innovation programme under grant agreement No. 101058715.
}

\author[First]{Christoph Nölle}
\author[First]{Petri Kannisto}

\address[First]{VDEh-Betriebsforschungsinstitut (BFI), \\ 
   Sohnstr. 69, 40237 Duesseldorf, Germany (e-mail: christoph.noelle\(\mid\)petri.kannisto@bfi.de).}

\begin{abstract} 
In the pursue for sustainability in process industry, digital twins necessitate the communication and storage of timeseries data about Industrial Internet of Things (IIoT).
Regarding timeseries, this paper first presents a set of requirements specific to process industries.
Then, it surveys how existing IIoT technologies meet the requirements.
The technologies include the API specifications Asset Administration Shell (AAS), Digital Twin Definition Language (DTDL), NGSI-LD and Open Platform Communications Unified Architecture (OPC UA) as well as six commercial platforms.
All the technologies leave significant gaps regarding the requirements, which means that tailor-made extensions are necessary.
\end{abstract}


\end{frontmatter}

\section{Introduction}
\input{10_intro}

\section{Related Work}
\label{sec:background}
\input{20_background}

\section{Requirements from Process Industry}
\label{sec:requirements}
\input{30_requirements}

\section{Comparison of Technologies}
\label{sec:main_result}
\input{40_comparison}

\section{Discussion}
\label{sec:discussion}
\input{50_discussion}

\section{Conclusions and Outlook}
\label{sec:concl}
\input{60_concl}

\bibliography{ifacconf}

\end{document}

%% file: 10_intro.tex
Timeseries data is an important ingredient in Industrial Internet of Things (IIoT) applications, and this is particularly true for process industries that strive towards sustainability goals, such as \mycite{eu_green_deal2050_2019}{2019}.
In sustainability efforts, digital twins (DT) are deployed to optimise production, reduce emissions and contribute to circular economy.
In DTs, one of the key challenges is interoperability in data exchange \citep{perno_et_al2022digi_twin_proc_ind_review}.
For DTs to operate, the exchanged data must cover long time horizons, which is possible only with timeseries structures.

Timeseries data require special treatment in an IoT platform, because they are usually stored in a dedicated timeseries database, offering columnar storage and compression.
Besides, special application programming interfaces (API) are necessary, providing among others filter options for the time range and aggregation functions.

In this paper, we present a collection of requirements on an IIoT platform that we extracted from the steel use case in the CAPRI research project \citep{noelle_et_al2022digi_twin_arch_proc_ind}. We further investigate in, how far several established IIoT platforms, as well as a set of API specifications, cover these requirements.
Therefore, the research objectives are:
\begin{enumerate}
    \item Recognize requirements for timeseries platforms in process industry, generalizing from the continuous casting process in steelmaking
    \item Survey IIoT platforms and API specifications to evaluate these against the requirements
\end{enumerate}

While many of the requirements are general in nature and should apply in other use cases too, some of our requirements are related to material tracking, which is somewhat specific to the process industries. The details of this case are explained in the following sections.

The remainder of this paper is organized as follows.
Section~\ref{sec:background} reviews the background.
Section~\ref{sec:requirements} presents the requirements recognized for timeseries, forming the basis of the technology survey and comparison in Section~\ref{sec:main_result}.
Finally, Section~\ref{sec:discussion} discusses the results and is followed by a conclusion in Section~\ref{sec:concl}.

%% file: 20_background.tex
Earlier articles have researched timeseries databases, but they omit the presentation of timeseries on IIoT platforms.
\cite{bader_et_al2017survey_open_src_time_series} compare 12 timeseries databases regarding multiple features, such as aggregate functions, sample resolution, APIs, scalability and load balancing.
\cite{petrik_et_al2019zeitreihendatenbanken} present a set of qualitative requirements for industrial timeseries databases, but these lack a focus on timeseries and rather consider more generic features, such as security, load management, analytics and storage base management.
\cite{mazumdar_et_al2019survey_data_storage_cloud_big_data} compare databases, including timeseries databases but not considering timeseries-specific features.
\cite{barez_et_al2023benchmarking_high_freq} benchmark the performance of timeseries databases for high-frequency data.
None of the discovered studies considers the functionality of timeseries presentation to an extent similar to this paper.
Furthermore, although the databases can offer timeseries functionality similar to IIoT platforms, IIoT widens the scope to actual devices.

Regarding IIoT platforms and technologies, earlier studies seem to focus on more generic questions.
These include, among others, interoperability \citep{kannisto_et_al2022plantwide_interop_msg_bus}, architecture, security and communications \citep{tan_and_samsudin2021tech_security_chall_iiot_survey} or security, digital twins and analytics support \citep{sauer_christian_2021_4485756}.
That is, this paper is situated in a research gap, offering a comparative study about the timeseries capabilities of IIoT platforms rather than researching either timeseries databases or IIoT platforms in general.

%% file: 30_requirements.tex
\subsection{Case Example from Steel Production}
\label{sec:sub:case}

Sensor data collected at a central IIoT platform a priori relate to time and form a timeseries. Every data point, providing a sensor measurement either as a single value or group of values, comes equipped with a timestamp, which uniquely identifies the data point. The set of data points over time forms a timeseries.

If material tracking is in place, our time-related sensor measurements can be assigned to physical positions in products or batches of material.
This assignment occurs by following the timestamps backwards in the production chain until the position of the sensor is reached. The sensor data are therefore not only related to a machine but also to products. 
If DTs are applied, it should be possible to refer to the data from both the machine twin and the product twins. Furthermore in the product perspective, the time-related aspect may be less relevant than the position-related aspect, and therefore data should be re-indexed by the position. 

\begin{figure}[tb]
    \begin{center}
    \includegraphics[scale=0.88]{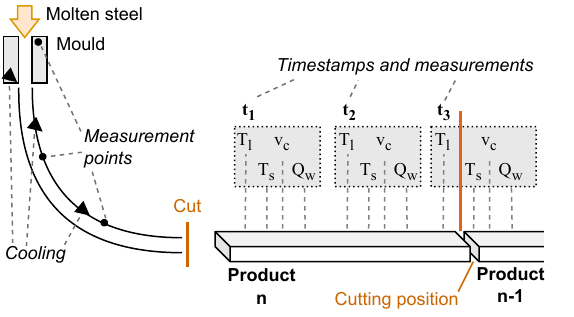}
    \caption{Schematic illustration of continuous casting and the mapping from time-indexed measurements to positions in the product.}
    \label{fig:cc}
    \end{center}
\end{figure}

For a concrete use case, let us take an example from the CAPRI project, considering the continuous casting process in steel production.
The casting machine consists of multiple bent strands, through which the initially liquid steel flows while cooling down and solidifying (see Fig.~\ref{fig:cc}). At the top of the strand, there is a water-cooled quadratic or rectangular tube, called the mould, in which an initial solid shell forms at the edge of the strand, enclosing the remaining liquid steel \citep{vynnycky2019continuous_casting}.
The shell ensures that the strand maintains its cross-section even below the mould. Spray water cooling is applied below the mould to this solid shell to support the cooling process. When the strand has completely solidified, intermediate steel products are cut off, called either billets, blooms or slabs, depending on the format.
In our use case, we are dealing with billets which are transformed into round bars in a later processing step.
For a product slice, it takes roughly 20 minutes from the mould to cutting.

In continuous casting, multiple sensors provide measurements along the strand, measuring among others the casting speed (\(v_c\) in Fig.~\ref{fig:cc}), liquid steel temperature (\(T_l\)), strand surface temperature (\(T_s\)), mould water flow or spray water flow (\(Q_w\)) and pressures. Typically, these measurements are recorded with a common timestamp, for instance in the Manufacturing Execution System (MES), so they form a single timeseries with multiple values. Since these measurements are taken at different positions, the values from an individual data point with a fixed timestamp belong to different positions on the end product, or even to different products. The assignment to products can only occur retrospectively when the exact cutting positions are known.

In order to convert the timeseries data to position-indexed
product data, it is convenient to record the total casted
length as an additional parameter in the timeseries.
This is done by integrating the casting speed over time (we assume
that the casting speed is constant along the strand).
For an individual billet,
the values for a sensor can be determined via an appropriate 
filter on the integrated length. 

Note that although this approach is sensor specific, by resampling 
the obtained values to equidistant points along the product length,
it is possible to obtain a new view on the original timeseries
as a multivariate table indexed by product id and product position. 
In this view, different sensor 
values at the same position may arise from different timestamps in
the original table. 

Instead of the product position, it is also possible to use an auxiliary
timestamp for indexing the new table, for instance the timestamp when 
casting started for the respective product slice. It is notable, however, that
sensor values taken further down the strand have actually
been retrieved with a time offset, therefore the time-based
view can be misleading. From a practical perspective, if
the underlying platform comes equipped with a dedicated
timeseries database, it may still be appropriate to use time
as the primary index for the derived table.

Note also that the derived table may be either stored explicitly in a new table or might instead be derived on demand, similarly to materialized and non-materialized views in traditional relational databases. Copying the data incurs the cost of additional storage but may speed up later queries. Additionally, mixed concepts could be a good compromise in some cases, storing for instance only the most recent data and generating the views on demand for older data.

How could an IIoT platform support the realization of such transformed views? A possible implementation option is:

\begin{enumerate}
    \item Enable access to the underlying timeseries database so that an admin user can create views on existing tables, as explained above, either materialized or not.
    \item The metamodel for the platform should support 1-dimensional data structures (sensor values over product length in our example). As a workaround, the data structure could be in the form of a timeseries again, as explained above.
    \item Allow the user to edit the connection between a field in the data model to a column in the database. 
\end{enumerate}

Such an approach could allow for very generic transformations, supporting also other use cases than the re-indexing to a length-based representation.

The approach does impose certain constraints on the timeseries database, however. For instance, resampling sensor values to equidistant points along a product can be difficult with the popular timeseries database InfluxDb, since it only supports GROUP BY operations on time and tags (columns of low cardinality)\footnote{\myurl{https://docs.influxdata.com/influxdb/cloud/query-data/influxql/explore-data/group-by/}}, but not on a \emph{length} field. Therefore, the mechanism described above would need to be adapted for this particular database, or the resampling would need to be implemented at platform level instead of a database view.

\subsection{IIoT Platform Requirements}
\label{sec:sub:iiot_platf_reqs}

Generalizing from continuous casting, the following requirements pertaining to timeseries handling should be supported by an IIoT platform.
To fulfil a particular requirement, some features are considered mandatory.
Some are merely preferable, annotated with the word \emph{ideally} in the following paragraphs.

\subsubsection{1. Basic support / API.}
The platform must be able to store timeseries data assigned to an asset (e.g., machine or product), and it must provide an API to retrieve the asset-related timeseries data. This API must define filters for time ranges and paging and \emph{ideally} aggregation functions.

\subsubsection{2. Streaming to clients.}
Timeseries data tend to become large quickly, so data streaming should be supported by the application-facing platform API. Note that even simple HTTP responses can be consumed in a streaming fashion, therefore no sophisticated protocols need to be implemented to support this requirement. However, the widely adopted JSON data format is problematic in this respect, because it requires the complete data to be transferred before the response can be validated (e.g., if there is a closing bracket). Alternatives could be JSON streaming\footnote{\myurl{https://en.wikipedia.org/wiki/JSON_streaming}} and/or a binary format such as Apache Arrow\footnote{\myurl{https://arrow.apache.org/}}. Instead of streaming, paging support can be seen as an alternative, but it requires more effort by the application developer, since a new request must to be sent after each batch of data points. Therefore, we consider paging support insufficient to fulfil this requirement.

\subsubsection{3. Multivariate timeseries.}
It should be possible to store multiple values as separate attributes into a single timeseries. Assume that a sensor, device or backend system reports multiple values with a single common timestamp similar to the steel example presented in Section~\ref{sec:sub:case}. Then, it should be possible to both store the values and to retrieve them in this way, instead of having to duplicate the timestamp for every individual series.

\subsubsection{4. Historical data auto-storage.} When a new value arrives for some attribute, then the platform should persist it into the historical data storage automatically if so configured. It should be possible to define the configuration settings for this feature at different levels: (1) globally, (2) per table/entity type or (3) per attribute, with more granular settings taking precedence over coarser ones.
From levels (1)-(3), at least two should be supported.

\subsubsection{5. Data retention and aggregation.}
It is usually infeasible to keep timeseries data around forever, in particular historical data. After a certain time, sometimes only an aggregated value is kept, such as hourly or daily values. Aggregated values are useful even before the data are cleaned up, for instance for analytics applications. Ideally, the platform should support the automated generation of aggregated timeseries and the automated cleanup of data after a configurable retention period. Both aggregation and retention features should support fine-grained settings, allowing the user to define them (1) globally, (2) per table/entity type and (3) per individual attribute, with more granular settings taking precedence over coarser ones. Among levels (1)-(3), at least two should be supported. Regarding query access to historic data, the platform should \emph{ideally} select the most appropriate aggregation level to build the response for a query. Suppose that the historic data of an attribute is persisted in raw format and additionally aggregated both at hourly and daily level. If a query requests for the weekly averages over a certain period, then the platform could automatically select the daily aggregation level for retrieving this data.

\subsubsection{6. Metadata for timeseries.}
It should be possible to provide metadata about a timeseries, both in a static way (the same for all entities of a given type) and individually per entity. Metadata can include semantic information about the timeseries (e.g., historical data vs. predictions vs. operating schedules), units used, etc. In the storage layer, it should not be replicated for every single data point, since this could lead to a dramatic increase in size. Sometimes metadata may only apply to a subset of data points, however, and \emph{ideally} there should be mechanisms to allow for the provision of segmented metadata.

\subsubsection{7. Beyond historical timeseries.}
Most timeseries data encountered in IoT use cases are historical, but there are other types, such as operating schedules or forecasts. These should be supported as well, and \emph{ideally} the metamodel of the platform should provide some standardized way to convey the meaning of a timeseries.

\subsubsection{8. Multiple references.}
A given timeseries may be relevant to multiple entities. For instance, it may be referred to by the DT of a machine and (part of it) by the DT of a product. In our steel casting example there may even be multiple products of different aggregation levels referring to the same timeseries, such as a bar, a billet and a heat, where the bar timeseries is a subset of the billet timeseries, and the latter is a subset of the heat timeseries. The platform should allow for different entities to refer to the subsets of a single timeseries.

\subsubsection{9. Transformation views.}
Different views into timeseries data should be supported, as explained in Section~\ref{sec:sub:case}, where an original time-indexed data stream is transformed into a length- or position-indexed data stream. Both materialized views (copying of data) and non-materialized views (on-demand generation) should be supported.

\subsubsection{10. Legacy data integration.}
Often a significant amount of historical data is already available in an existing database or in files. It should be possible to integrate these data into the platform without physical copying. Due to the heterogeneity of possible data sources, a plugin interface for software extensions could be a suitable option.
Additionally, the platform would \emph{ideally} support some commonly used timeseries formats out of the box.
Another aspect of such legacy or external datastores is that timeseries would \emph{ideally} be composable from different sources. For instance, one datastore could provide the data points before a certain timestamp and another one the newer data points.

%% file: 40_comparison.tex
\subsection{Selection Criteria and Survey Method}

We investigated selected technologies, including APIs specifications and proprietary platforms, as to whether they support the above listed requirements.
Among APIs, the ones included must have a wide adoption either within industrial production systems or the related research.
Respectively, the proprietary platforms must have been ranked either \emph{leaders}, \emph{challengers} or \emph{visionaries} but not \emph{niche players} in \mycite{gartner_magic_2022}{2022} by Gartner.

The investigation is merely based on the documentation publicly available.
Thus, no experiments take place.

\subsection{Technologies}

The surveyed APIs are 4 in total.
Asset Administration Shell or AAS \mycitep{aas_idta_02008_1_1_2023}{2023} is an effort to develop a standard for digital asset representations, or digital twins, focusing on manufacturing use cases.
Digital Twins Definition Language \mycitep{dtdl_v3_2023}{2023} is an API specification from Microsoft to exchange data with DTs.
\mycite{ngsi_ld_2023}{2023} is an API and information model, based on linked data and developed as an open standard. It has emerged in the context of FIWARE initiative, and implementations are also known as FIWARE context brokers. Open Platform Communications Unified Architecture \mycitep{opc_ua_p1_2022}{2022} is a communication framework and open standard for the interoperability of systems and equipment.

Respectively, the included proprietary technologies are 6 in total:
\mycite{ability_2019}{2019} by ABB;
\mycite{aws_iot_sitewise_2022}{2022} and \mycite{aws_iot_twinmaker_2021}{2021} by Amazon;
\mycite{azure_iot_nd}{n.d.} and Azure Digital Twins by Microsoft;
\mycite{cumulocity_iot_nd}{n.d.} by Software AG;
\mycite{insights_hub_2023}{2023}, formerly MindSphere, by Siemens; and
\mycite{thingworx_2023}{2023} by PTC.
Some of these exploit the earlier mentioned APIs: Ability provides an OPC UA server, whereas Azure tools implement DTDL.
Although classified in \mycite{gartner_magic_2022}{2022}, Hitachi Vantara Lumada was excluded due to the unavailability of public documentation.
Amazon has an alternative platform, AWS Timestream, but SiteWise and TwinMaker appear more focused on IIoT.

\input{49_comparison_table}

Regarding ABB Ability, the review focuses on Ability History despite the newer product Ability Symphony Plus Historian.
Unfortunately, the latter seems to have no public documentation, but it is assumed to offer at least the same features as the former, which provides a public documentation from 2019.

Azure provides a significantly larger functionality than included in DTDL, so both are treated separately. DTDL focuses on the data model only, whereas the other
API specifications included prescribe an API for interacting with the platform and thus have a considerably wider scope. Because it is nevertheless interesting to evaluate DTDL, the analysis includes it.

Not all OPC UA implementations need all features of the standard, and thus there is a set of profiles.
Thanks to the profiles, each implementation can focus on its core areas while preserving interoperability.
In this survey however, all OPC UA features are included regardless of the profiles.

\subsection{Survey Results}
\label{sec:sub:survey_results}

Table~\ref{table:comparison} shows the results of the comparison.
Because the comparison is based on the documentation, there can be inaccuracies.
Especially if there is no indication whether a feature is fully implemented, it is assumed not to exist.
Partially satisfied features are marked with '\mypartial{}'.

1. All technologies provide an API to retrieve values from timeseries.
However, AAS provides a non-normative API with only start and end time as the parameters.
DTDL only has a data stream to communicate with.

2. Regarding streaming to clients, only Cumulocity IoT provides a support, which is JSON streaming.
Most of the other technologies specify a REST API with no reference to streaming.
Some support paging though.

3. Although a logically straightforward feature, multivariate timeseries are supported only in four technologies.
Among these, only AAS is a standard.

4. Historical data auto-storage is a basic feature, but not all platforms reached the configurability requirements.
NGSI-LD leaves the configuration features to implementers, and OPC UA lacks the levels of granularity.
InsightsHub and ThingWorx lack configurability.
Both AAS and Cumulocity lack a reference to this feature altogether.

5. Data retention and aggregation are powerful for not only saving space but also increasing the performance of data retrieval.
Here, the API specifications perform poorly, only OPC UA providing aggregation features but still lacking retention.
The platforms perform better, but Azure lacks aggregation and InsightsHub configuration respectively.

6. Not all technologies allow metadata for an entire timeseries.
In DTDL, the feature is out of scope, and Insights Hub provides no information about the feature.
In NGSI-LD and Cumulocity, the metadata can seemingly be stored per timestamp only.
The others enable at least timeseries-level metadata.
AAS excels in this regard, enabling metadata even for timeseries segments.

7. Beyond historical timeseries means that the timestamps can be in the future.
Unfortunately, Ability and AWS SiteWise categorically prevent this. 
In DTDL, timeseries are foreseen only for historization.

8., 9. The advanced features \emph{Multiple references} and \emph{Transformation views} receive almost no match.
AAS does not specify the mapping to the storage layer, therefore implementations can in principle support these features. Still, they are out of scope of the standard. 

OPC UA enables a value (along with its history) to be referred to from multiple nodes, but this enables no time ranges.
In practice, any use case with these requirements necessitates data duplication.

10. The platforms vary in the support for external timeseries storages.
NGSI-LD, Ability and AWS TwinMaker fulfil the requirement.
NGSI-LD provides zero-copy integration for legacy data via a plugin interface, which still requires engineering efforts.
Similarly, both Ability and AWS TwinMaker enable custom connectors.
AAS can store a query to linked sources, but the client would still execute the query by itself, which is a partial match. This feature, supporting even segmentation of the timeseries into different parts, is a feature unseen in the other technologies and very useful from the point of view of legacy data integration. The design decision to expose this segmentation to the data consumer appears more questionable, though.

The rest are considered no support.
An OPC UA server can aggregate others, but the specification provides no support for non-OPC timeseries formats.
ThingWorx supports external connectivity, but it is unclear if or how this applies to querying external timeseries sources.
The other documentations lack a reference to the feature.

Overall, no technology is even close to cover all the requirements.
No commercial platform covers more than a half of the requirements in full.
In comparison, the API specifications perform even lower on average.

%% file: 49_comparison_table.tex
\begin{table*}[hbt]

\begin{center}
\caption{Comparison of IIoT technologies with regard to the timeseries requirements.}\label{table:comparison}
\renewcommand{\arraystretch}{1.1} 
\begin{tabular}{p{4.5cm}|llll|llllll}
&
\multicolumn{4}{l|}{\textbf{Open standards, API specs}} & 
\multicolumn{6}{l}{\textbf{Proprietary platforms}} \\
&
AAS &
DTDL &
\multicolumn{1}{p{0.9cm}}{NGSI-LD} &
\multicolumn{1}{p{0.8cm}|}{OPC UA} &
Ability &
AWS &
Azure &
\multicolumn{1}{p{1.3cm}}{Cumulocity IoT} &
\multicolumn{1}{p{0.85cm}}{Insights Hub} &
\multicolumn{1}{p{0.8cm}}{Thing-Worx} \\
\hline
1. Basic support / API &
\mypartial{} &
\mypartial{} &
\myyes{} &
\myyes{} &
\myyes{} &
\myyes{} &
\myyes{} &
\myyes{} &
\myyes{} &
\myyes{} \\
2. Streaming to clients &
\myno{} &
\myno{} &
\myno{} &
\myno{} &
\myno{} &
\myno{} &
\myno{} &
\myyes{} &
\myno{} &
\myno{} \\
3. Multivariate timeseries &
\myyes{} &
\myno{} &
\myno{} &
\myno{} &
\myno{} &
\myno{} &
\myno{} &
\myyes{} &
\myyes{} &
\myyes{} \\
4. Historical data auto-storage &
\myno{} &
\myyes{} &
\mypartial{} &
\mypartial{} &
\myyes{} &
\myyes{} &
\myyes{} &
\myno{} &
\mypartial{} &
\mypartial{} \\
5. Data retention and aggregation &
\myno{} &
\myno{} &
\myno{} &
\mypartial{} &
\myyes{} &
\myyes{} &
\mypartial{} &
\myyes{} &
\mypartial{} &
\myyes{} \\
6. Metadata for timeseries &
\myyes{} &
\myno{} &
\myno{} &
\myyes{} &
\myyes{} &
\myyes{} &
\myyes{} &
\myno{} &
\myno{} &
\myyes{} \\
7. Beyond historical timeseries &
\myyes{} &
\myno{} &
\myyes{} &
\myyes{} &
\myno{} &
\myno{} &
\myyes{} &
\myyes{} &
\myyes{} &
\myyes{} \\
8. Multiple references &
\mypartial{} &
\myno{} &
\myno{} &
\mypartial{} &
\myno{} &
\myno{} &
\myno{} &
\myno{} &
\myno{} &
\myno{} \\
9. Transformation views &
\myno{} &
\myno{} &
\myno{} &
\myno{} &
\myno{} &
\myno{} &
\myno{} &
\myno{} &
\myno{} &
\myno{} \\
10. Legacy data integration &
\mypartial{} &
\myno{} &
\myyes{} &
\myno{} &
\myyes{} &
\myyes{} &
\myno{} &
\myno{} &
\myno{} &
\myno{} \\
\hline
\end{tabular}
\end{center}
\end{table*}

%% file: 50_discussion.tex

Because all of the technologies surveyed exhibit significant gaps (Table~\ref{table:comparison}) regarding the requirements (Subsection~\ref{sec:sub:iiot_platf_reqs}), software solutions require workarounds, for instance in the form of custom modules, manual integration steps or data duplication.
This applies to any non-trivial timeseries handling in IIoT platforms and DTs. 

The technological gaps can be seen as an opportunity for the open initiatives to define the standards for handling of timeseries data in IIoT platforms. At the moment, however, the proprietary platforms appear generally more complete and consistent.

The requirements collected are sufficiently generic to be applicable in other use cases, as well, but their relative importance has to be determined case by case. Although DTs are applied in many domains and use cases, process industry applications are particularly important from a sustainability point of view due to their heavy energy usage, relatively large environmental footprint and societal importance regarding the products.

The main limitations of this work come firstly from the inductively derived requirements, focusing on timeseries, and secondly from the documentation-based survey.
More requirements could be recognized in the broader context of digital twins, including aspects such as linked data, knowledge graphs, query language, multimedia data, etc. 
We highlight the need for further research in this area. 
Additionally, the survey builds upon public documentation.
Where the documentation is incomplete, imprecise or complex, misinterpretation is possible. Even when the documentation is clear, it is not always straightforward to decide whether each requirement is fulfilled in a satisfying way.
Besides, ABB Ability was evaluated based on an older product due to the lack of public documentation for the most recent.

%% file: 60_concl.tex
We have presented a set of requirements pertaining to timeseries handling in IIoT platforms, from the point of view of process industries. Additionally, several commercial platforms and open standards have been investigated with regards to their support for these requirements, and significant gaps have been identified. The platforms are evolving dynamically, however, with new services and capabilities being announced frequently, so that the fulfilment matrix is bound to change as well.
This has an effect on how the platforms and APIs serve digital twins for sustainable industrial production.

What we have described here for timeseries data, i.e. one-dimensional sensor measurements over time, applies in similar form to other types of data, for instance image streams/videos from optical inspection systems. They can be relevant both from a machine and product perspective as well, and they often need to be transformed from a time-based to position-based view but with two- or three-dimensional coordinates. The requirements for efficient data storage, indexing and APIs are even more profound in this case. \citep{brandenburger_2016_hr_server}

Digital twin platforms represent not only entities and their properties but also relationships between different entities, forming a graph. Several platforms offer support for some form of graph query languages and corresponding visualizations, and an extension of our analysis to linked data capabilities could be an interesting pursuit.

Another interesting question is whether standardization efforts for timeseries APIs could be relevant to IIoT platforms. For instance, under the umbrella of the Apache Arrow project, the FlightSQL standard for querying of columnar datastores has been specified.\footnote{\myurl{https://arrow.apache.org/docs/format/FlightSql.html}} It has been adopted among others by the InfluxDB database. Similarly, the PostgreSQL wire protocol\footnote{\myurl{https://www.postgresql.org/docs/current/protocol.html}} used by TimescaleDB and some others, such as QuestDB and CrateDB, can be seen as a standardized protocol. These standards could simplify the integration of external data stores, and could possibly also provide the basis for efficient binary timeseries APIs offered by the platforms.

%% file: 00_main.bbl
\begin{thebibliography}{24}
\providecommand{\natexlab}[1]{#1}
\providecommand{\url}[1]{\texttt{#1}}
\providecommand{\urlprefix}{URL }
\expandafter\ifx\csname urlstyle\endcsname\relax
  \providecommand{\doi}[1]{doi:\discretionary{}{}{}#1}\else
  \providecommand{\doi}{doi:\discretionary{}{}{}\begingroup
  \urlstyle{rm}\Url}\fi

\bibitem[{Ability History()}]{ability_2019}
Ability History (2019).
\newblock ABB.
\newblock \url{https://docs.cpmplus.net/docs} [Visited 20 Sep 2023]. Version
  5.2.

\bibitem[{AWS IoT SiteWise()}]{aws_iot_sitewise_2022}
AWS IoT SiteWise (2022).
\newblock Amazon.
\newblock \url{https://docs.aws.amazon.com/iot-sitewise/latest/userguide}
  [Visited 20 Sep 2023]. Version 2019-12-02.

\bibitem[{AWS IoT TwinMaker()}]{aws_iot_twinmaker_2021}
AWS IoT TwinMaker (2021).
\newblock Amazon.
\newblock \url{https://docs.aws.amazon.com/iot-twinmaker/latest/guide} [Visited
  20 Sep 2023]. Version 2021-11-30.

\bibitem[{Azure IoT Hub()}]{azure_iot_nd}
Azure IoT Hub (n.d.).
\newblock Microsoft.
\newblock \url{https://azure.microsoft.com/en-us/products/iot-hub/} [Visited 31
  Aug 2023]. Unknown version.

\bibitem[{Bader et~al.(2017)Bader, Kopp, and
  Falkenthal}]{bader_et_al2017survey_open_src_time_series}
Bader, A., Kopp, O., and Falkenthal, M. (2017).
\newblock Survey and comparison of open source time series databases.
\newblock In \emph{Datenbanksysteme für Business, Technologie und Web (BTW
  2017) - Workshopband}, 249--268.

\bibitem[{Barez et~al.(2023)Barez, Bilokon, and
  Xiong}]{barez_et_al2023benchmarking_high_freq}
Barez, F., Bilokon, P., and Xiong, R. (2023).
\newblock Benchmarking specialized databases for high-frequency data.
\newblock \url{https://doi.org/10.48550/arXiv.2301.12561} [Visited 26 Sep
  2023].

\bibitem[{Brandenburger et~al.(2016)Brandenburger, Colla, Nastasi, Ferro,
  Schirm, and Melcher}]{brandenburger_2016_hr_server}
Brandenburger, J., Colla, V., Nastasi, G., Ferro, F., Schirm, C., and Melcher,
  J. (2016).
\newblock Big data solution for quality monitoring and improvement on flat
  steel production.
\newblock \emph{IFAC-PapersOnLine}, 49, 55--60.
\newblock \doi{10.1016/j.ifacol.2016.10.096}.

\bibitem[{Cumulocity IoT()}]{cumulocity_iot_nd}
Cumulocity IoT (n.d.).
\newblock Software AG.
\newblock \url{https://cumulocity.com/guides/concepts/introduction/} [Visited
  20 Sep 2023]. Version 10.17.0.

\bibitem[{DTDL()}]{dtdl_v3_2023}
DTDL (2023).
\newblock Digital twins definition language.
\newblock
  \url{https://azure.github.io/opendigitaltwins-dtdl/DTDL/v3/DTDL.v3.html}
  [Visited 31 Aug 2023]. Version 3.

\bibitem[{{E}uropean Green Deal()}]{eu_green_deal2050_2019}
{E}uropean Green Deal (2019).
\newblock {COM}(2019) 640 final.
\newblock European Commission.
\newblock
  \url{https://eur-lex.europa.eu/legal-content/EN/TXT/PDF/?uri=CELEX:52019DC0640}
  [Visited 23 Aug 2023].

\bibitem[{Global Industrial IoT Platforms()}]{gartner_magic_2022}
Global Industrial IoT Platforms (2022).
\newblock Gartner magic quadrant.
\newblock \url{https://www.gartner.com/en/documents/4022099} [Visited 29 Aug
  2023].

\bibitem[{IDTA 02008-1-1()}]{aas_idta_02008_1_1_2023}
IDTA 02008-1-1 (2023).
\newblock Time series data.
\newblock
  \url{https://industrialdigitaltwin.org/wp-content/uploads/2023/03/IDTA-02008-1-1_Submodel_TimeSeriesData.pdf}
  [Visited 15 Sep 2023]. Version 1.1.

\bibitem[{Insights Hub()}]{insights_hub_2023}
Insights Hub (2023).
\newblock Siemens.
\newblock \url{https://plm.sw.siemens.com/en-US/insights-hub/} [Visited 20 Sep
  2023]. Version 2023-09-16 (Europe 1).

\bibitem[{Kannisto et~al.(2022)Kannisto, Hästbacka, Gutiérrez, Suominen,
  Vilkko, and Craamer}]{kannisto_et_al2022plantwide_interop_msg_bus}
Kannisto, P., Hästbacka, D., Gutiérrez, T., Suominen, O., Vilkko, M., and
  Craamer, P. (2022).
\newblock Plant-wide interoperability and decoupled, data-driven process
  control with message bus communication.
\newblock \emph{Journal of Industrial Information Integration}, 26, 100253.
\newblock \doi{10.1016/j.jii.2021.100253}.

\bibitem[{Mazumdar et~al.(2019)Mazumdar, Seybold, Kritikos, and
  Verginadis}]{mazumdar_et_al2019survey_data_storage_cloud_big_data}
Mazumdar, S., Seybold, D., Kritikos, K., and Verginadis, Y. (2019).
\newblock A survey on data storage and placement methodologies for cloud-big
  data ecosystem.
\newblock \emph{Journal of Big Data}, 6(15).
\newblock \doi{10.1186/s40537-019-0178-3}.

\bibitem[{NGSI-LD()}]{ngsi_ld_2023}
NGSI-LD (2023).
\newblock {ETSI} {GS} {CIM} 009 {NGSI}-{LD} {API} {V}1.7.1.
\newblock
  \url{https://www.etsi.org/deliver/etsi_gs/CIM/001_099/009/01.07.01_60/gs_CIM009v010701p.pdf}.
\newblock [Visited 30 Aug 2023].

\bibitem[{Nölle et~al.(2022)Nölle, Arteaga, Egia, Salis, {De Luca}, and
  Holzknecht}]{noelle_et_al2022digi_twin_arch_proc_ind}
Nölle, C., Arteaga, A., Egia, J., Salis, A., {De Luca}, G., and Holzknecht, N.
  (2022).
\newblock Digital twin-enabled application architecture for the process
  industry.
\newblock In \emph{Proceedings of the 3rd International Conference on
  Innovative Intelligent Industrial Production and Logistics - ETCIIM},
  255--266. INSTICC, SciTePress.
\newblock \doi{10.5220/0011561800003329}.

\bibitem[{OPC UA()}]{opc_ua_p1_2022}
OPC UA (2022).
\newblock Part 1: Overview and concepts, release 1.05.02.
\newblock OPC Foundation.

\bibitem[{Perno et~al.(2022)Perno, Hvam, and
  Haug}]{perno_et_al2022digi_twin_proc_ind_review}
Perno, M., Hvam, L., and Haug, A. (2022).
\newblock Implementation of digital twins in the process industry: A systematic
  literature review of enablers and barriers.
\newblock \emph{Computers in Industry}, 134, 103558.
\newblock \doi{10.1016/j.compind.2021.103558}.

\bibitem[{Petrik et~al.(2019)Petrik, Mormul, and
  Reimann}]{petrik_et_al2019zeitreihendatenbanken}
Petrik, D., Mormul, M., and Reimann, P. (2019).
\newblock Anforderungen für {Z}eitreihendatenbanken in der industriellen
  {E}dge.
\newblock \emph{HMD Praxis der Wirtschaftsinformatik}, 56, 1282--1308.
\newblock \doi{10.1365/s40702-019-00568-9}.

\bibitem[{Sauer et~al.(2021)Sauer, Eichelberger, Ahmadian, Dewes, and
  Jürjens}]{sauer_christian_2021_4485756}
Sauer, C., Eichelberger, H., Ahmadian, A.S., Dewes, A., and Jürjens, J.
  (2021).
\newblock Current {I}ndustry 4.0 platforms – an overview.
\newblock \doi{10.5281/zenodo.4485756}.

\bibitem[{Tan and
  Samsudin(2021)}]{tan_and_samsudin2021tech_security_chall_iiot_survey}
Tan, S.F. and Samsudin, A. (2021).
\newblock Recent technologies, security countermeasure and ongoing challenges
  of industrial internet of things ({II}o{T}): A survey.
\newblock \emph{Sensors}, 21(19).
\newblock \doi{10.3390/s21196647}.

\bibitem[{ThingWorx()}]{thingworx_2023}
ThingWorx (2023).
\newblock PTC.
\newblock \url{https://support.ptc.com/help/thingworx/platform/r9/en} [Visited
  20 Sep 2023]. Version 9.4.1.

\bibitem[{Vynnycky(2019)}]{vynnycky2019continuous_casting}
Vynnycky, M. (2019).
\newblock Continuous casting.
\newblock \emph{Metals}, 9(6).
\newblock \doi{10.3390/met9060643}.

\end{thebibliography}
